\documentclass[9pt,a4paper,notitlepage,twocolumn,oneside]{article}

\title{Demixing, remixing and cellular networks in binary liquids containing colloidal particles}
\author{Job H.~J.~Thijssen$^{\mathrm{*a,b}}$ and Paul S.~Clegg$^{\mathrm{a,b}}$}

\usepackage[sort&compress,numbers,super]{natbib}
\usepackage{mciteplus}
\usepackage[dvips]{graphicx}
\usepackage{titleref}

\newcommand{\degC}[0]{$^{\circ} \mathrm{C}$}
\newcommand{\degCpm}[0]{$^{\circ} \mathrm{C} \cdot \mathrm{min}^{-1}$}
\newcommand{\micron}[0]{$\mu$m}

\begin{document}

\maketitle

\noindent{\footnotesize{\textit{$^{\mathrm{a}}$SUPA School of Physics and Astronomy, The University of Edinburgh, Mayfield Road, Edinburgh EH9 3JZ, United Kingdom. E-mail: j.h.j.thijssen@ed.ac.uk}}}

\noindent{\footnotesize{\textit{$^{\mathrm{b}}$COSMIC, The University of Edinburgh, Mayfield Road, Edinburgh EH9 3JZ, United Kingdom.}}}


\begin{abstract}
We present a confocal-microscopy study of demixing and remixing in binary liquids containing colloidal particles. First, particle-stabilized emulsions have been fabricated by nucleation and growth of droplets upon cooling from the single-fluid phase. We show that their stability mainly derives from interfacial particles; the surplus of colloids in the continuous phase possibly provides additional stability. Upon heating these emulsions, we have observed the formation of polyhedral cellular networks of colloids, just before the system remixes. Given a suitable liquid-liquid composition, the initial emulsions cross the binary-liquid symmetry line due to creaming. Therefore, upon heating, the droplets do not shrink and they remain closely packed. The subsequent network formation relies on a delicate balance between the Laplace pressure and the pressure due to creaming/remixing. As high concentrations of colloids in the cell walls inhibit film thinning and rupture, the networks can be stabilized for more than 30 minutes. This opens up an avenue for their application in the fabrication of advanced materials.
\end{abstract}


\section{Introduction}

Colloidal particles in composites involving immiscible liquids are encountered in many products and processes, including food emulsions and crude-oil extraction.\cite{BinksHorozov2008} Using partially miscible liquids, phase separation instead of direct mixing can be used as a means to organize the particles.\cite{Clegg2007} The complex interplay between particle wettability and liquid-liquid demixing has been studied extensively and has been shown to lead to wetting-induced interactions,\cite{Hertlein2008,Araki2008,Lu2008,Koehler1997,Beysens1999,Gallagher1992} a percolation transition\cite{Loewen1995,Peng2000} and arrested states such as particle-stabilized emulsions and bijels. \cite{Clegg2007,Stratford2005,Herzig2007} In all of these examples, the presence of a dispersed constituent affects a (reversible) phase transition into an \emph{ordered} state. Though it has been shown that thermal cycling in particle-stabilized emulsions can lead to extended fluid domains,\cite{Clegg2007} the effects of solid particles on a (reversible) phase transition into a \emph{disordered} state are virtually unexplored. In this paper, we consider both demixing and remixing in binary liquids containing colloidal particles and we report the formation of polyhedral cellular networks of colloids en route to remixing.

\begin{figure*}
        \begin{center}
        \includegraphics[width=0.5\textwidth]{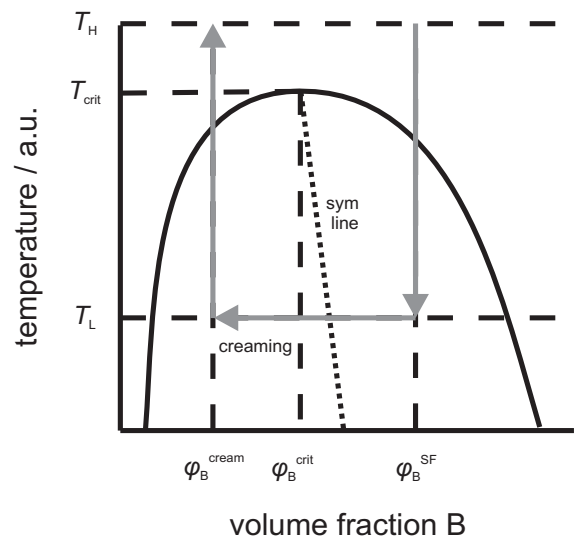}
            \caption{Schematic phase diagram of a binary mixture of liquids A and B. An emulsion is formed from the single-fluid phase at a temperature $T_{\mathrm{H}}$ by quenching to $T_{\mathrm{L}}$. Due to creaming, the volume fraction of the continuous-phase liquid (B) locally decreases from $\varphi_{\mathrm{B}}^{\mathrm{SF}}$ to $\varphi_{\mathrm{B}}^{\mathrm{cream}}$; the system thus crosses the symmetry line. To study remixing, the creamed emulsion is heated from $T_{\mathrm{L}}$ to $T_{\mathrm{H}}$. (See Sec.~\ref{SIpart:NfComp} for system-specific phase diagrams.)}\label{fig:Figure_Phase_Diagram}
        \end{center}
    \end{figure*}

When a binary liquid is quenched into the two-phase region of its phase diagram, there are several ways in which the liquid components can demix.\cite{Tanaka2000} If the phase diagram is symmetric and the sample is of critical composition (see $\varphi_{\mathrm{B}}^{\mathrm{crit}}$ in Fig.~\ref{fig:Figure_Phase_Diagram}), a deep quench will induce spinodal decomposition, resulting in a bicontinuous domain pattern. Here, neither of the phases is the dispersed phase, so the line of temperatures at this composition can be called the static symmetry line (Fig.~\ref{fig:Figure_Phase_Diagram}). If the composition is off-critical (e.g.~$\varphi_{\mathrm{B}}^{\mathrm{SF}}$ in Fig.~\ref{fig:Figure_Phase_Diagram}), and the quench is not too deep, liquid-liquid demixing proceeds via nucleation and growth of droplets. Now there is a dispersed phase and the symmetry has been broken. A further degree of complexity emerges if the two phases that are separating have very different viscoelastic characteristics.\cite{Tanaka2000}

In most emulsions, a third component is added to prevent complete separation of the two liquid phases. In conventional emulsions, droplets are stabilized by the addition of a molecular surfactant; about a century ago, however, it was realized that solid particles can be used as emulsifiers instead.\cite{BinksHorozov2008} They can strongly attach to liquid-liquid interfaces, for they reduce energy-expensive liquid-liquid contact area. This is described by
\begin{equation}\label{Eq:Uattach}
    \Delta G_{\mathrm{d}} = \pi r^{2} \gamma \left( 1 - \left| \cos{\theta} \right| \right)^{2} \ ,
\end{equation}
where $\Delta G_{\mathrm{d}}$ is the free energy of detachment of a spherical particle of radius $r$ and contact angle $\theta$, as measured through the more polar phase.\cite{BinksHorozov2008} Even for the partially miscible liquids hexane and methanol, $\left| \Delta G_{\mathrm{d}} \right| \sim 6 \cdot 10^{4} \ k_{\mathrm{B}} T_\mathrm{room}$ for an `average' colloidal particle, i.e.~1 \micron{} in diameter.\cite{Abbas1997,HandbookChemPhys20082009} Thus, unlike molecular surfactants, particles can become permanently stuck at liquid-liquid interfaces.

Particles need not be at the interface between the phases to affect their separation. For example, the formation of particle networks in the continuous phase can kinetically stabilize emulsions.\cite{BinksHorozov2008} In polymer blends, partitioning of solid (nano)particles into one of the phases has been shown to slow down domain growth and alter pattern formation.\cite{Tanaka1994,Chung2004} Geometric constraints due to the partitioning give rise to particle assemblies with high packing fractions in one of the phases, typically the dispersed or minority phase, changing coarsening dynamics by spatial and/or shape-pinning. Finally, in foams, there are a host of mechanisms through which non-adsorbing particles can provide stability, including reduced film thinning/rupture due to stratification.\cite{Hunter2008}

Here, we present polyhedral cellular networks of colloids formed by heating particle-stabilized emulsions into the single-fluid phase. The structures, which we will simply refer to as `cellular networks', resemble liquid-liquid foams and the observed excess of particles in the continuous phase may well play a key role. The initial particle-stabilized emulsions can be formed by nucleation and growth of droplets upon cooling from the single-fluid phase. Both liquid-liquid demixing and remixing in the presence of colloidal particles are studied using confocal microscopy, including the effects of binary-liquid composition, particle wettability, colloid volume fraction and heating rate. Employing a binary liquid with a symmetric phase diagram, we have fabricated nearly space-filling networks. As these can be stabilized for more than 30 minutes, they may find applications as templates for advanced materials.\cite{Banhart2001,Cook2008}

The rest of this paper is organized as follows. In Sec.~\ref{Sec:Materials_and_methods}, we describe the experimental procedures, including sample preparation, cooling and heating protocols, and characterization techniques. In Sec.~\ref{Sec:Results_and_Disc}, we present the formation of particle-stabilized emulsions through demixing and the formation of cellular networks en route to remixing. We then discuss the formation of the observed networks in Sec.~\ref{Sec:discussion}, where we also put our results in a broader context. Finally, in Sec.~\ref{Sec:Conclusions}, we summarize our findings and draw conclusions.


\section{Materials and methods}\label{Sec:Materials_and_methods}

\subsection{Sample preparation}\label{subsec:Sample_preparation}

The majority of the experiments were performed with silica colloids in the binary mixture hexane-methanol (Fig.~\ref{SIfig:Figure_SI_PhaseDiagrams} in Sec.~\ref{SIpart:NfComp}).\cite{Hradetzky1991} Hydrophobic, fumed silica particles were used as received (Degussa, AEROSIL R812). The average primary particle size was $\sim 7 \ \mathrm{nm}$ and they were roughly spherical (supplier data and Transmission Electron Microscopy (TEM)). Effects of particle wettability were investigated using a range of roughly spherical, fumed silica particles (Wacker-Chemie) with an average diameter $\sim 15$ nm (TEM). These were kindly provided by B.~Binks and had been silanized such that 32\% (FS32), 42\% (FS42) or 62\% (FS62) of the (hydrophilic) silanol groups remained on their surface. Hexane (Fluka, $\ge 99$\%), methanol (Fisher Scientific, $99.99$\%) and the fluorescent dye Nile Red (Sigma) were used as received.\cite{WaterNote,Alessi1989} For liquid-liquid composition and particle-wettability sweeps, iso-hexane (Rathburn, RH1002) and methanol (VWR, $\ge 99.8\%$) were used instead.\cite{IsohexaneNote} Nile Red was usually dissolved in the methanol at concentrations of $3.7 \cdot 10^{-4}$ to $2.9 \cdot 10^{-3}$ M. Sample mixtures were typically prepared at a hexane/methanol volume ratio of 65/35 (61/39 w/w) and a silica content of 2.0 vol-\% (5.4 wt-\%). The particles were dispersed using an ultrasonic processor at 6 W for 2 minutes (Sonics, Vibra-cell), with part of the vial immersed in water, followed by 10 s of vortex mixing.

To test the generality of our results, a few experiments were performed using poly(methyl methacrylate) (PMMA) particles in the binary mixture cyclohexene-nitromethane (Fig.~\ref{SIfig:Figure_SI_PhaseDiagrams} in Sec.~\ref{SIpart:NfComp}).\cite{Sazonov2000} The PMMA particles had been synthesized following Bosma \textit{et al}\cite{Bosma2002}, after which they had been washed in pentane. Their cores were labeled with the fluorescent dye 4-chloro-7-nitrobenzo-2-oxa-1,3-diazol (NBD). According to Static Light Scattering (SLS), they had an outer diameter of 1.06 \micron{} and a polydispersity of 7\%. Cyclohexene (Acros Organics, 99\%) and nitromethane (Acros Organics, 99+\%) were used as received. Nile Red was dissolved in the nitromethane at a concentration of $8.5 \cdot 10^{-4}$ M. Sample mixtures were typically prepared at a cyclohexene/nitromethane volume ratio of 63/37 (55/45 w/w) and a PMMA volume fraction of 2.1\% (2.6 wt-\%). The dried particles were dispersed using the ultrasonic processor at 6 W for $3 \times 1$ minute.

All amounts were determined by weighing. Evaporation losses during sonication led to a maximum deviation of 2.5\%-point and 3.5\%-point in the hexane/methanol and cyclohexene/nitromethane volume ratios, respectively. Regular blank tests, without colloidal particles and/or without Nile Red, were performed to check for residual stabilization.

\subsection{Transfer, cooling and heating}\label{subsec:Transfer_cooling_and_heating}

Sample cells were rectangular and had a 1 mm internal path length (Starna Scientific Ltd). Hexane-methanol/silica mixtures in the single-fluid phase were transferred to these cells in an incubator at approximately 40 \degC{} (Stuart, SI60). The samples were then quickly transferred to a modified hotstage at 40.0 \degC{} (Linkam Scientific LTS350). Particle-stabilized emulsions were most usually formed by cooling from 40.0 \degC{} to 22.0 \degC{} at 5.0 \degCpm{} (Fig.~\ref{fig:Figure_Phase_Diagram}). Subsequently, remixing was studied by warming from 22.0 \degC{} to 40.0 or 45.0 \degC{} at selected rates between 1.0 \degCpm{} and 20.0 \degCpm{}. Typically, the time span between cooling and warming was 5 to 25 minutes. During an experimental session, samples were usually subjected to several heating and cooling cycles. After a heating cycle, they were vigorously shaken by hand in the incubator at 40 \degC{} to ensure proper particle re-dispersion.

In a few cases, we deviated from the standard procedure described above. Firstly, for liquid-liquid composition and particle wettability sweeps, mixtures were transferred by Pasteur pipette immediately after they had reached the single-fluid regime due to sonication. Secondly, samples with less than 1.0 vol-\% of silica and those for two-photon fluorescence microscopy were quenched from $\sim 40$ \degC{} by putting them on a metal surface at $\sim 22$ \degC{}. Finally, during two-photon fluorescence experiments, sample temperature was controlled using a LakeShore 331 controller connected to an in-house manufactured hotstage. In this case, typical heating rates were $\sim 0.1$ \degCpm{}.

Cyclohexene-nitromethane/PMMA emulsions were formed by 15 s of vortex mixing at room temperature. The resulting emulsions were transferred to the sample cells by Pasteur pipette. This procedure resulted in higher volume fractions of nitromethane in the sample cells. Throughout this paper, cyclohexene-nitromethane ratios are those before transfer. Samples were heated from 22.0 \degC{} to 53.0 \degC{} at 20.0 \degCpm{}.

\subsection{Sample characterization}\label{subsec:sample_char}

\subsubsection{Confocal microscopy}\label{subsec:Confocal_microscopy}

The samples were studied with confocal laser scanning microscopy in reflection, fluorescence and/or transmission. A Nikon ECLIPSE E800/TE300 upright/inverted microscope was used in conjunction with a BioRad Radiance 2100 (MP) scanning system. The 457 nm line of an Ar-ion laser was used for reflection, while the 488 nm line was used to excite NBD in PMMA and Nile Red in nitromethane. A 543 nm HeNe laser was employed to excite Nile Red in methanol. Filters were used as appropriate. Owing to the hotstage, Nikon Plan Fluor Extra Long Working Distance objectives with an adjustable correction collar were used: (1) $20\times$/0.45 NA and (2) $60\times$/0.70 NA.

Visual inspection and confocal microscopy on samples without particles confirmed that the recorded Nile Red fluorescence was mainly coming from the methanol-rich and nitromethane-rich phases. However, the cyclohexene-rich phase may appear brighter in confocal fluorescence images than the nitromethane-rich phase. This is due to contrast inversion, as confirmed by 3D confocal scans. Note, in addition, that the Nile Red fluorescence contrast at room temperature is significantly smaller in the cyclohexene-nitromethane system than it is in the hexane-methanol system.

\subsubsection{Two-photon fluorescence microscopy}\label{subsec:TwoPhoton_fluorescence_microscopy}

Two-photon fluorescence microscopy was performed using a pulsed 76 MHz, mode-locked laser (Coherent, Mira 900), tuned to approximately 790 nm (OceanOptics, USB2000) and operated at an average output power of 34 mW (Spectra-Physics, Model 407A). To allow for multiphoton excitation in hexane-methanol/silica samples, the fluorescent dye 8-Anilino-1-naphthalenesulfonic acid (ANS, Fluka, $\ge 97$\%) was added to the methanol at a concentration of $5.1 \cdot 10^{-4}~\mathrm{M}$. A 625 nm short-pass filter was used to separate reflection and emission.

To obtain $xz$-images, $xyz$-series were resliced halfway along the $y$-axis using the ImageJ software package,\cite{ImageJ138x} thus obtaining $xzy$-series. Spanning $\sim 6$ \micron{}, 11 consecutive frames in the $xzy$-series were averaged. The $z$-axis was rescaled by measuring the internal path length of a cuvette, filled with a hexane-methanol/silica mixture in the single-fluid phase, using confocal microscopy.


\section{Results}\label{Sec:Results_and_Disc}


\subsection{Emulsions}\label{Sec:resultsdemixing}

We employed confocal microscopy to study the structure of hexane-methanol/silica emulsions that had been formed by quenching from the single-fluid phase (see \emph{Supplementary Information} for movies). Figs.~\ref{fig:Figure_Network_Formation}a and \ref{fig:Figure_Network_Formation}e show the reflection and the corresponding fluorescence images. The fluorescence indicates that the droplets contain the hexane-rich phase. This observation is consistent with both the creaming behavior and the phase diagram - the hexane-rich phase is less dense\cite{Abbas1997} and it is the minority phase.\cite{Hradetzky1991} The confocal reflection image reveals that the silica particles mainly partition into the continuous phase. Some colloids may be trapped on the hexane-methanol interface, but the resolution of our confocal images is not sufficient to establish that.

To establish the emulsion-stabilization mechanism, we prepared hexane-methanol/silica emulsions at colloid volume fractions as low as 0.1\%. These formed via macroscopically non-uniform nucleation, which seems to aid the droplets in sweeping up particles. The confocal reflection image in Fig.~\ref{fig:Figure_3D_Stab}a shows bright lines around the droplets, consistent with interfacially trapped particles and little surplus in the continuous phase.\cite{BlankNote} Moreover, at such low silica volume fractions, it seems unlikely that colloids in the continuous phase solely can stabilize our emulsions for at least 40 minutes. Thus, our hexane-methanol/silica emulsions are stabilized by interfacial particles, leaving an excess of colloids in the continuous phase.

In order to test the dependence on particle wettability, we prepared emulsions with particle batches of systematically controlled hydrophobicity. Silica particles FS32 yield hexane-in-methanol emulsions that are qualitatively similar to those obtained with the default R812 silica. A stable emulsion was also formed with slightly more hydrophilic silica particles (FS42). However, the colloid volume fraction had to be decreased to 1.3\%, for the sample could not be properly mixed at 2.5 vol-\%. Finally, using hydrophilic silica (FS62), emulsion preparation was even more challenging. Hence, the emulsification we observe is associated with a small range of particle wettabilities.

We examined the relationship between emulsification and phase behavior by varying the liquid-liquid composition. In between 70/30 and 50/50 v/v, our particle-stabilized emulsions are qualitatively similar to those at the default ratio of 65/35 v/v - on the phase diagram, the system is just pushed along the methanol-rich branch (Fig.~\ref{SIfig:Figure_SI_PhaseDiagrams} in Sec.~\ref{SIpart:NfComp}). Decreasing the methanol content below 30 vol-\% takes the system toward the hexane-rich branch. Confocal fluorescence reveals that the system inverts at 80/20 v/v, forming methanol-in-hexane emulsions, which do not cream to the top of the cuvette. Unfortunately, it is not clear from confocal reflection images where the silica resides in these inverted emulsions.

To summarize, we have shown that demixing in hexane-methanol mixtures containing nanoscale, hydrophobic silica particles is a viable route for the fabrication of particle-stabilized emulsions. Apart from particle wettability, the effects of liquid-liquid composition have been investigated and can be explained using the phase behavior of the binary liquid.


\subsection{Cellular networks}\label{Sec:resultsremixing}

\subsubsection{Network formation}\label{subsubsec:Network_formation}

Our emulsions, prepared from partially miscible liquids, are stable on experimental timescales (Figs.~\ref{fig:Figure_Network_Formation}a/e). Upon warming, the emulsions coarsen, mainly due to coalescence; the area fraction of the dispersed phase is seen to increase. Further heating leads to the formation of a polyhedral cellular network (Figs.~\ref{fig:Figure_Network_Formation}b/c) just before the system remixes (Fig.~\ref{fig:Figure_Network_Formation}d). We have found networks in both hexane-methanol/silica and cyclohexene-nitromethane/PMMA; the two systems are presented together in Fig.~\ref{fig:Figure_Network_Formation}. From these image sequences, two features are clear: (1) the reflection(silica)/NBD(PMMA) contrast increases upon network formation, suggesting that the colloids become more densely packed; (2) the Nile Red fluorescence contrast decreases significantly, indicating that the two liquid phases approach one another in chemical composition. The resulting networks resemble a liquid-liquid foam (Figs.~\ref{fig:Figure_Network_Formation}c/k).

After observing network formation in 2D, we turned to two-photon fluorescence microscopy to probe the 3D structure of our hexane-methanol/silica samples. Fig.~\ref{fig:Figure_TwoPhoton} shows vertical ($xz$) and horizontal ($xy$) images of such a system just below and just above the network-formation temperature.\cite{AveragingNote} As demonstrated in Figs.~\ref{fig:Figure_TwoPhoton}a/b, samples could be imaged clearly up to a depth of four droplet layers. Heating the emulsion from 35.5 \degC{} to 36.0 \degC{}, a cellular network formed near the top of the cuvette (Figs.~\ref{fig:Figure_TwoPhoton}c/d). Note that these networks are thicker than a single layer. Interestingly, the droplets/cells in the bottom layer of the cream, where there are no droplets pushing from below, remained (semi-)spherical. Upon further heating (Figs.~\ref{fig:Figure_TwoPhoton}e/f), the network coarsened, the bottom half of the lowest cells still remaining curved.

Intriguingly, Fig.~\ref{fig:Figure_TwoPhoton} implies that networks form above the consolute temperature of the hexane-methanol mixture.\cite{Hradetzky1991,Abbas1997,Alessi1989} To check whether the presence of colloids affects the phase behavior of the binary liquid, we determined the nucleation temperature of our hexane-methanol mixtures in a 0.2 \degCpm{} quench. Yielding values of 33.3 \degC{} for a sample without colloids and 35.8 \degC{} for a sample containing 2.6 vol-\% of silica, this suggests that the particles raise the binodal at this particular hexane-methanol volume ratio by approximately 1 \degC{}/vol-\%.

Employing the cyclohexene-nitromethane/PMMA system, we have been able to fabricate nearly space-filling networks (see also Sec.~\ref{SIpart:NfComp}). As it turns out, their full 3D structure can be imaged using standard confocal microscopy (Fig.~\ref{fig:Figure_3D_Stab}b). This shows us that, throughout the height of the sample, walls between a small and a large cell bulge into the larger one, which agrees with the Laplace pressure being inversely proportional to cell size. Fig.~\ref{fig:Figure_3D_Stab}c implies that some of the colloidal particles do end up on the interface of the cells and that they form closely packed structures in the continuous `backbone' of the cellular network.

Networks in both systems can be temporarily stabilized at a temperature close to the binodal (Figs.~\ref{fig:Figure_3D_Stab}(d-f)). However, comparing Figs.~\ref{fig:Figure_3D_Stab}e with \ref{fig:Figure_3D_Stab}f, it is clear that the average cell size slowly increases at constant temperature on a timescale of 33 minutes. The time-evolution of the system bears a remarkable resemblance to the dynamics of foams. Comparable features include creaming, T1 re-arrangements and cell coalescence. Given these similarities, we attribute the slow coarsening of our networks at constant temperature to drainage of liquid and particles from the films between droplets. Upon further heating, the networks do fall apart.

\subsubsection{Parameter tuning}\label{subsubsec:Network_parameters}

To better understand and tune network formation, we now explore the effects of varying liquid-liquid composition, particle wettability, colloid volume fraction and heating rate in the hexane-methanol/silica system. This system was chosen because the samples can be easily rejuvenated after each study.

Regarding liquid-liquid composition, we have found that networks were only formed for hexane/methanol ratios between 60/40 and 70/30 v/v. To the right (50/50 v/v), which is further from the critical composition (Fig.~\ref{SIfig:Figure_SI_PhaseDiagrams} in Sec.~\ref{SIpart:NfComp}),\cite{Hradetzky1991,Abbas1997,Alessi1989} the droplets shrink and disappear upon warming. To the left (80/20 v/v), methanol-in-hexane emulsions form upon cooling from the single-fluid phase, indicating that the system meets the hexane-rich branch only.

As emulsification depends strongly on particle wettability (Sec.~\ref{Sec:resultsdemixing}), one would expect the same to be true for network formation. Using FS32 instead of R812 silica, we have fabricated qualitatively similar networks. At half the typical colloid volume fraction of 2.5\%, we have also been able to obtain a network-like structure using the more hydrophilic FS42 particles. However, this formed more slowly, its cells appeared to be less densely packed and it took the fluorescence contrast longer to fade. Using hydrophilic silica (FS62), emulsion droplets just shrank and disappeared upon warming. (Note that initial emulsification was challenging with these particles.)

To tune the porosity of our cellular networks, we systematically varied colloid volume fraction and heating rate. Figs.~\ref{fig:Figure_ColloidPhi_HeatingRate}(a-d) reveal the effects of increasing the particle content at a fixed heating rate of 3.0 \degCpm{}. Initially, the average cell size in the network just decreases, indicating that coarsening through coalescence is suppressed. However, adding even more silica causes the walls of the network to become diffuse and finally the droplets just disappear, without forming facetted cells. Alternatively, increasing the heating rate at a fixed colloid volume fraction of 2.0\% also causes the average cell size of the networks to decrease (Figs.~\ref{fig:Figure_ColloidPhi_HeatingRate}(a,e-g)). We have observed that, for each colloid volume fraction, there is a minimum warming rate below which the cell size is of the order the field of view and a maximum warming rate above which the droplets just disappear, without forming facetted cells.

\subsubsection{Summary}\label{subsubsec:Network_summary}

We have shown that cellular networks of colloids can be fabricated by heating particle-stabilized emulsions into the single-fluid phase. These networks will only form if the dispersed phase (droplets) is no longer the minority phase. Network formation may occur at temperatures above the binodal, at which the liquid phases are similar in composition, but the surplus of colloidal particles in the continuous phase prevents them from fully remixing. At these temperatures, we have been able to stabilize nearly space-filling networks for more than 30 minutes, their time-evolution at constant temperature resembling that of conventional foams. We have shown that the formation of networks is sensitive to liquid-liquid composition and particle wettability. In addition, their morphology can be tuned by varying colloid volume fraction and/or warming rate.


\section{Discussion}\label{Sec:discussion}

\subsection{Network formation and stability}\label{subsec:network_mechanism}

\begin{figure*}
        \begin{center}
            \includegraphics[width=1.0\textwidth]{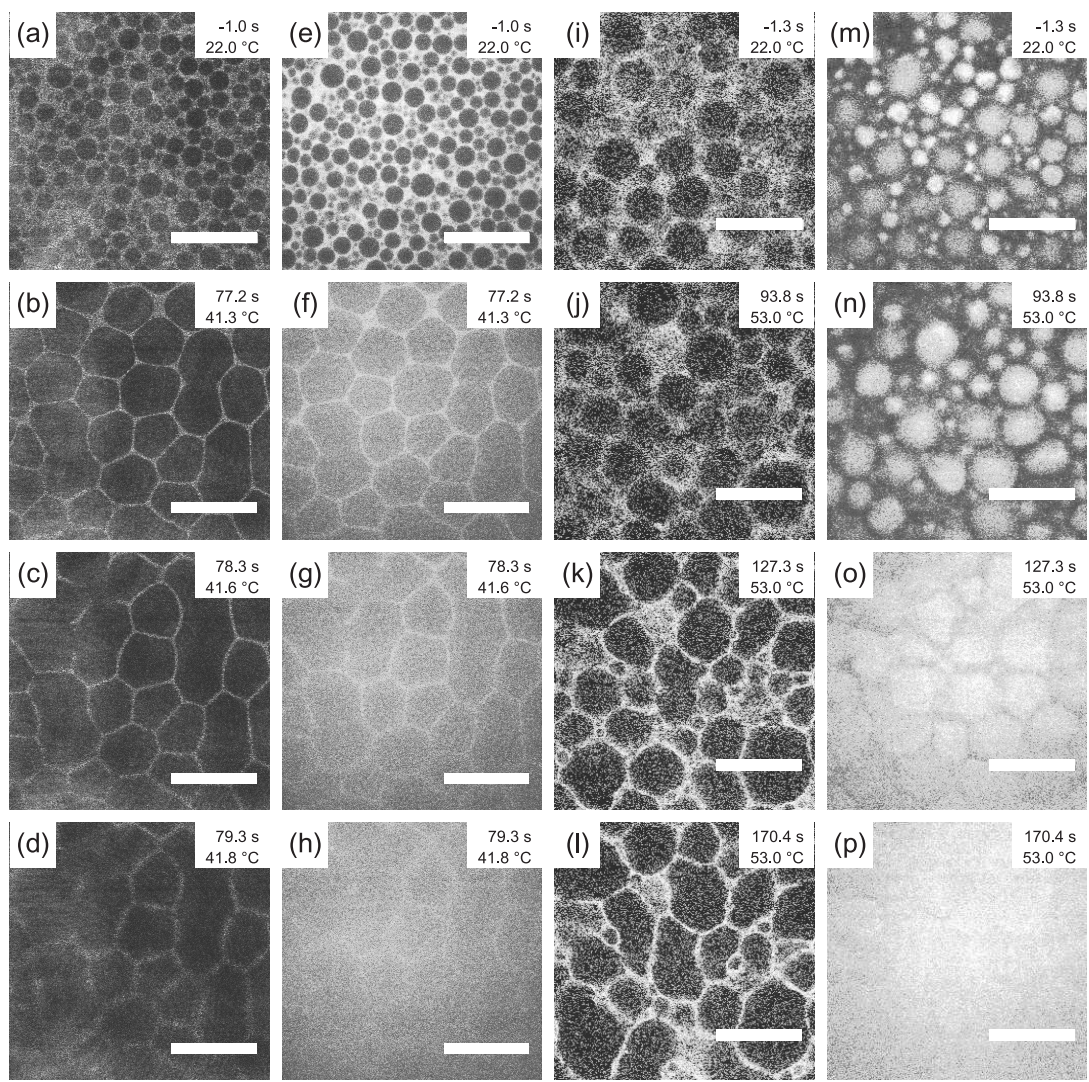}
            \caption{Top to bottom: selected frames from confocal $xyt$-series showing the formation of cellular networks from (a-h) hexane-methanol/silica and (i-p) cyclohexene-nitromethane/PMMA emulsions. Starting at $t = 0$ s, the sample was heated (a-h) from 22.0 \degC{} to 45.0 \degC{} at 15.0 \degCpm{} and (i-p) from 22.0 \degC{} to 53.0 \degC{} at 20.0 \degCpm{}. Left to right: the columns display reflection (silica), Nile Red fluorescence (methanol), NBD fluorescence (PMMA) and Nile Red fluorescence (cyclohexene, see Sec.~\ref{subsec:Confocal_microscopy}). Images were taken at (a-h) 14 \micron{} and (i-p) 30 \micron{} from the top of the sample. Scale bars: 100 \micron{}. (See \emph{Supplementary Information} for movies.)}\label{fig:Figure_Network_Formation}
        \end{center}
    \end{figure*}

\begin{figure*}
        \begin{center}
            \includegraphics[width=1.0\textwidth]{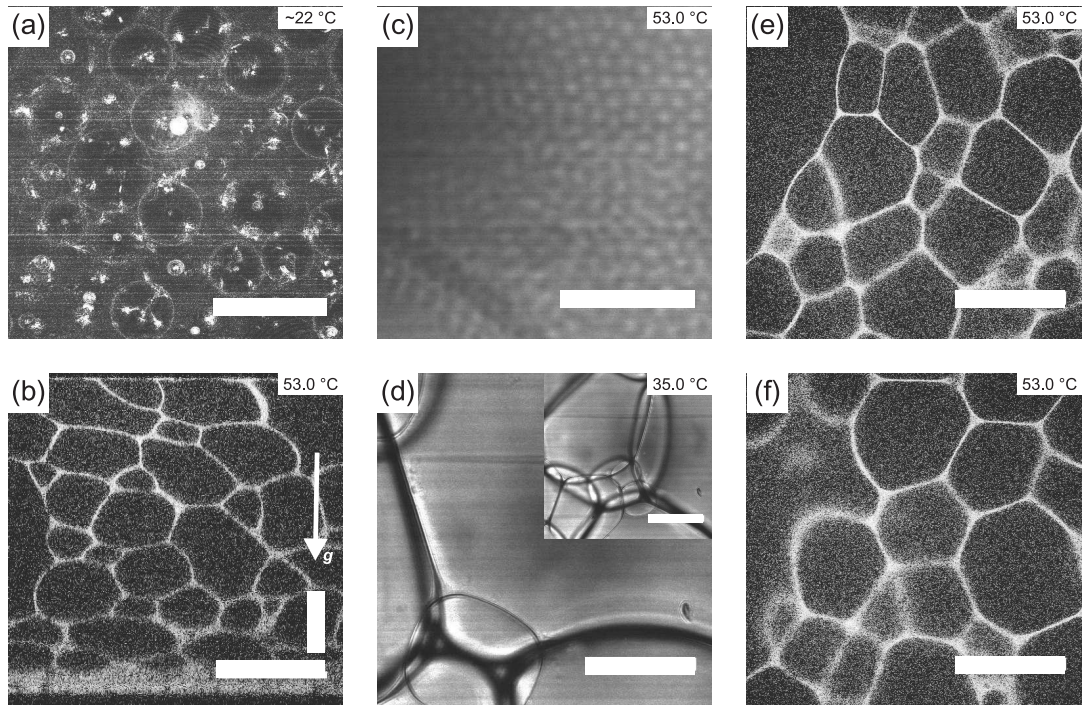}
            \caption{(a) Confocal reflection $xy$-image of a hexane-methanol/silica emulsion (0.1 vol-\% silica), showing interfacially trapped particles. (b) Confocal fluorescence $xz$ (vertical) and (c) transmission $xy$ (horizontal) images of colloids in a cyclohexene-nitromethane/PMMA network held at 53.0 \degC{}. (d) Transmission $xy$-image of a hexane-methanol/silica network held at 35.0 \degC{}; the inset shows the same network $\sim 8$ minutes before. (e,f) Confocal fluorescence $xy$-images, $\sim 33$ minutes apart, of colloids in a cyclohexene-nitromethane/PMMA network held at 53.0 \degC{} (see \emph{Supplementary Information} for movie). Images were recorded at a depth of (a) 0.79 mm from the bottom of the sample and (c) 0.11 mm, (d) 8 \micron{} and (e,f) 0.44 mm from the top of the sample. Scale bars: (a) 100 \micron{}, (b) 200 \micron{}, (c) 10 \micron{}, (d) 100 \micron{} and (e,f) 200 \micron{}.}\label{fig:Figure_3D_Stab}
        \end{center}
    \end{figure*}

\begin{figure*}
        \begin{center}
            \includegraphics[width=1.0\textwidth]{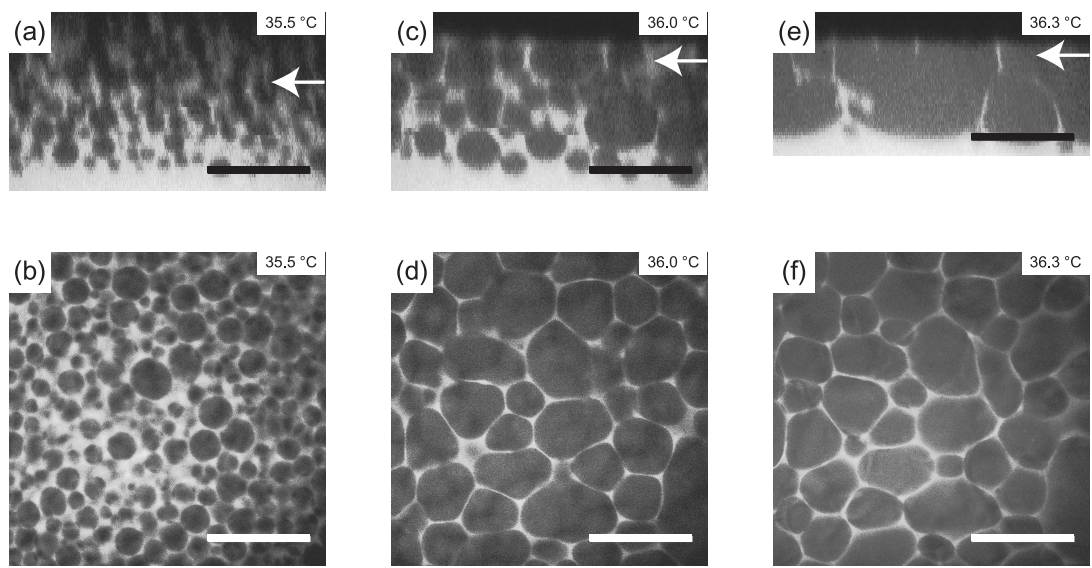}
            \caption{3D structure from two-photon fluorescence images of the formation of a cellular network by step-wise heating of a creamed hexane-methanol/silica emulsion: (a,c,e) vertical $xz$-slices and (b,d,f) horizontal $xy$-slices; the latter were taken at the heights indicated in panels (a,c,e) by white arrows (see Sec.~\ref{subsec:TwoPhoton_fluorescence_microscopy} for details)\cite{AveragingNote}. Sample temperature was constant for $\ge 3$ minutes prior to imaging; heating rates were $\sim 0.1$ \degCpm{}. Scale bars: 200 \micron{}.}\label{fig:Figure_TwoPhoton}
        \end{center}
    \end{figure*}

Our proposed mechanism for network formation is shown schematically in Fig.~\ref{fig:Figure_Network_Schematic}. It relies on a delicate interplay between phase behavior, buoyancy and interfacial tension. The initial emulsions are stabilized by colloidal particles at the droplet interfaces, though additional stability may be derived from the excess of colloids in the continuous phase (Figs.~\ref{fig:Figure_Network_Schematic}a, \ref{fig:Figure_3D_Stab}a and \ref{fig:Figure_Network_Formation}). During emulsion preparation, these droplets nucleated as the minority phase when the sample was cooled/mixed. We observe that the droplets grow rather than shrink upon heating: this is the key feature driving the formation of a network. Droplet growth implies that the system must have crossed the symmetry line of the phase diagram. In both systems, this is (locally) accomplished by creaming (Fig.~\ref{fig:Figure_Phase_Diagram}), which also keeps the droplets closely packed.

Warming continuously toward the network-formation temperature, the chemical compositions of the two liquid phases become more alike (Fig.~\ref{fig:Figure_Network_Schematic}b), as expressed by the decreasing Nile Red fluorescence and refractive-index contrasts (Figs.~\ref{fig:Figure_TwoPhoton} and \ref{fig:Figure_Network_Formation}). The corresponding fall in interfacial tension lowers the energy barrier for the formation of fluid bridges between droplets, e.g.~by particle detachment or by reducing attractive capillary forces between interfacial particles.\cite{Hunter2008} This facilitates emulsion coarsening through coalescence (Figs.~\ref{fig:Figure_Network_Schematic}b and \ref{fig:Figure_Network_Formation}).

Not only does the decrease in interfacial tension facilitate coarsening, it also allows for droplet deformation. Far below the network-formation temperature, the Laplace pressure
\begin{equation}\label{Eq:LaplacePressure}
    p_{\mathrm{L}} = \frac{2 \cdot \gamma}{R}
\end{equation}
ensures that droplets remain spherical. Balancing Laplace and buoyancy-induced normal pressures at the contact area between the cuvette wall and a bare droplet, the ratio of the disc and droplet radii $\left( r / R \right)$ is
\begin{equation}\label{Eq:LaplaceBuoyancyBalance}
    \alpha = \left( \frac{r}{R} \right) = R \cdot \sqrt{\left( \frac{2 g}{3} \right) \cdot \left( \frac{\rho_{\mathrm{c}} - \rho_{\mathrm{d}}}{\gamma} \right)} \ .
\end{equation}
Here, $\rho_{\mathrm{c/d}}$ is the density of the continuous/dispersed phase and $g$ is the acceleration of free fall. Upon heating, the interfacial tension decreases faster than the density difference.\cite{Abbas1997} In the case of hexane-methanol, for example, this causes $\alpha \left( R = 15 \ \mu\mathrm{m} \right)$ to increase from $\sim 0.022$ at 25 \degC{} to $\sim 0.042$ at 32 \degC{},\cite{Abbas1997,HandbookChemPhys20082009} which is still a few \degC{} below the consolute temperature.\cite{Hradetzky1991,Abbas1997,Alessi1989} In addition, the droplet radius $R$ grows upon heating, due to coalescence, leading to a further increase in $\alpha$ and thus demonstrating that significant droplet deformation is possible close to the binodal. The diffusion of liquid material from the continuous phase to the dispersed phase, due to remixing, will only aid droplet deformation.

\begin{figure*}
        \begin{center}
            \includegraphics[width=0.5\textwidth]{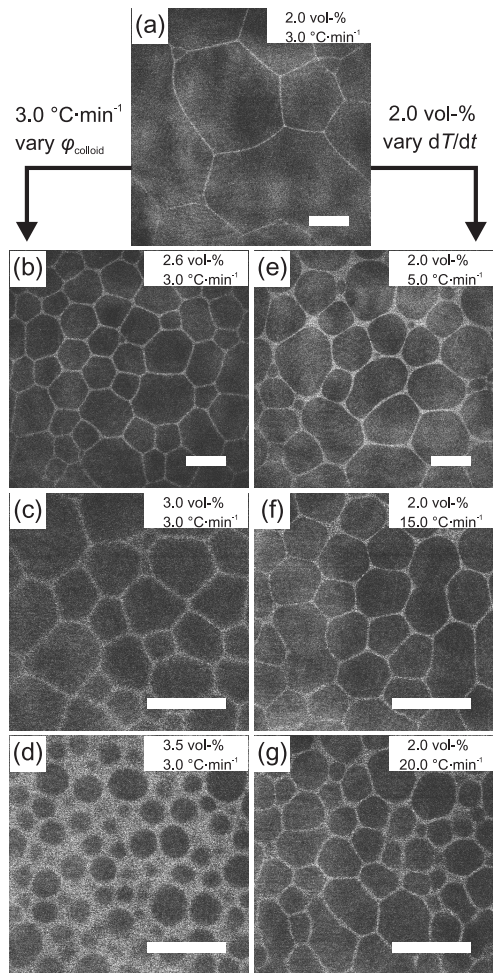}
            \caption{Reflection confocal images of silica particles (white) revealing the effects of varying (a-d) colloid volume fraction and (a,e-g) heating rate on cellular networks formed from hexane-methanol/silica emulsions. Images were taken at $\sim$14 \micron{} from the top of the sample at intervals of (a-d) 4.0 s, (e) 3.0 s and (f,g) 1.0 s. Scale bars: 100 \micron{}.}\label{fig:Figure_ColloidPhi_HeatingRate}
        \end{center}
    \end{figure*}

\begin{figure*}
        \begin{center}
            \includegraphics[width=1.0\textwidth]{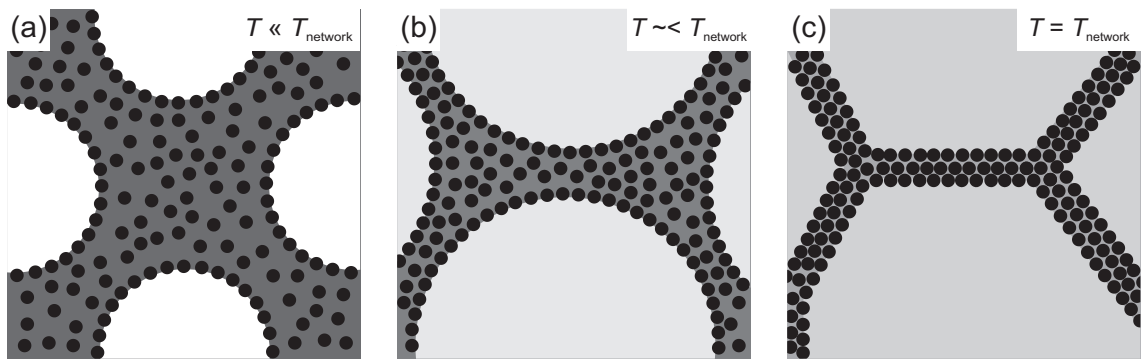}
            \caption{Schematic representation of the formation of a cellular network of colloids (see Sec.~\ref{subsec:network_mechanism} details). (a) Emulsion droplets (white) are stabilized by colloidal particles (black), a substantial proportion of which are dispersed throughout the continuous phase (dark gray). (b) Upon heating, the chemical compositions of the two liquid phases become more alike, the system coarsens and the number density of colloids in the continuous phase locally increases. (c) Approaching the network-formation temperature, the Laplace pressure can no longer balance the pressure due to creaming/remixing and the droplets deform. Consequently, the particles get pushed into a polyhedral cellular network.}\label{fig:Figure_Network_Schematic}
        \end{center}
    \end{figure*}

Since the particles are on the interface and in the minority phase, which is slowly disappearing, they are pushed into a cellular network as the droplets deform (Figs.~\ref{fig:Figure_Network_Schematic}c and \ref{fig:Figure_Network_Formation}). Indeed, the reflection/NBD fluorescence images in Fig.~\ref{fig:Figure_Network_Formation} and the high-magnification image in Fig.~\ref{fig:Figure_3D_Stab}c suggest that the cellular networks are stabilized by high concentrations of colloidal particles in the liquid films between the droplets, inhibiting film thinning and rupture.\cite{Hunter2008} This also agrees with the slow coarsening of our cellular networks at constant temperature (Figs.~\ref{fig:Figure_3D_Stab}(d-f)), which can be attributed to drainage of liquid and particles. Moreover, parameters that promote interfacial attachment or continuous phase viscosity, such as particle hydrophilicity and particle volume fraction, tend to impede both coalescence and network formation (Fig.~\ref{fig:Figure_ColloidPhi_HeatingRate}(a-d)). This corroborates the idea that dense packings of particles stabilize the cellular networks.

As remixing progresses, the particles in the network walls may retain an adsorbed/wetting layer of the continuous phase. This can lead to capillary necks and hence attractions between particles.\cite{Araki2008,Beysens1999,Gallagher1992,Hunter2008} Such attractive interparticle forces can increase colloid packing fraction and network stability, thus reducing film thinning and rupture. They could even lead to particle aggregation, allowing the particulate network to survive complete liquid-liquid remixing. Indeed, though networks do fall apart upon further heating, we have observed cell-wall remnants. Such network `resilience' may increase with time spent in the arrested state, an ageing process that has recently been observed and modeled in bijels.\cite{Sanz2009}

In common with almost all particle-stabilized emulsions, our networks are arrested out of equilibrium. Our structures are prepared from particle-stabilized droplet emulsions by heating them toward the single-fluid phase. Network formation is a robust feature of this process in that either quasi-stationary (Fig.~\ref{fig:Figure_TwoPhoton}) or fast temperature ramps (Figs.~\ref{fig:Figure_ColloidPhi_HeatingRate} and \ref{fig:Figure_Network_Formation}) can be employed. However, in all cases, the system is taken toward the single-fluid phase. Numerous researchers, including ourselves, have studied the behavior of particles dispersed in these and other binary liquids upon demixing.\cite{Clegg2007,Beysens1999,Koehler1997,Gallagher1992,Lu2008} None have observed the formation of polyhedral cellular networks as they approached the binodal. Such sensitivity to the preparation route, combined with interfacial trapping of particles, implies that the network is not an equilibrium structure.

\subsection{Context}\label{subsec:discussion_context}

It is illuminating to compare our cellular networks with those formed via other generic routes. Networks are often formed via the interaction between a dispersed constituent (particles, polymers, etc.) and a host solvent with an ordering transition (binary liquid, liquid crystal, etc.). For example, Tanaka \emph{et al.}~reported pattern evolution in phase-separating polymer blends in the presence of glass beads.\cite{Tanaka1994} The authors showed that spatial and shape-pinning effects significantly modify the coarsening dynamics of domains and they remark that these effects should be universal in any binary-liquid mixture containing solid particles. Subsequent studies of particles and polymers in phase-separating liquids have resulted in networks of these dispersed constituents, trapped at the liquid-liquid interface.

Similar networks have been seen in completely different systems. Meeker \emph{et al}.~reported the formation of self-supporting networks in mixtures of thermotropic liquid crystals and PMMA colloids upon cooling past the isotropic-to-nematic phase transition.\cite{Meeker2000} These liquid-crystal/particle networks are remarkably similar to the ones reported here.\cite{Meeker2000,Vollmer2005} As a final example, Colard \emph{et al.}~demonstrated that conducting nanocomposite polymer foams can be fabricated by freeze-drying a mixture of colloids in water, the newly formed ice crystals acting as a template for the porous structure.\cite{Colard2009}

In all of the above-mentioned examples, the system arrests in a reversible transition from a disordered state (mixed/isotropic/liquid) to an ordered state (demixed/nematic/crystal), via the pinning of defects. The colloid networks presented here were also formed by dispersing particles in a system with a reversible phase transition, but there the similarity ends. The networks were formed by beginning in the ordered phase (two-fluid) and warming toward the disordered phase (single-fluid). The networks in our case are not networks of defects in an ordered phase - pinned in place by a dispersed component. Instead, the networks appear to be held together by a wetting layer on the particles - a remnant of the ordered phase. Indeed, the interplay between wetting and phase separation remains full of surprises.


\section{Conclusions}\label{Sec:Conclusions}

In this paper, we have presented a confocal-microscopy study of liquid-liquid demixing and remixing in binary liquids containing colloidal particles. Particle-stabilized emulsions have been fabricated by nucleation and growth of droplets upon cooling from the single-fluid phase. Subsequently heating such emulsions, we have observed the formation of polyhedral cellular networks of colloids, just before the system remixes.

Regarding emulsification, we have found that our hexane-methanol/silica emulsions are stabilized by interfacial particles, the surplus of colloids in the continuous phase possibly providing additional stability. We have also demonstrated that emulsion formation is associated with a small range of particle wettabilities and that it is related to the phase behavior of the binary liquid.

Concerning cellular networks, we have observed their formation in two different binary-liquid/particle systems. Given a suitable liquid-liquid composition, the initial emulsions cross the symmetry line of the binary-liquid phase diagram due to creaming, which prevents the droplets from shrinking upon warming and keeps them closely packed. We propose that network formation itself relies on a delicate balance between the Laplace pressure and the pressure due to creaming/remixing. High concentrations of colloidal particles in the liquid films between droplets inhibit film thinning and rupture, possibly assisted by wetting-induced interparticle attractions. The networks can be stabilized at temperatures close to the binodal; our results imply that these temperatures are above the binodal of the binary liquid, which we attribute to the particles modifying the underlying phase behavior. Employing a binary liquid with a symmetric phase diagram, we have even stabilized nearly space-filling networks. Finally, network formation and morphology have been shown to depend on particle wettability, colloid volume fraction and heating rate.

In short, our results demonstrate that the interplay between (reversible) phase transitions and an additional dispersed component is not limited to systems en route to an \emph{ordered} state, but can significantly affect transitions into a \emph{disordered} state as well. From the point of view of applications, 3D cellular networks of colloids that are stable for more than 30 minutes may be used as templates in the fabrication of advanced materials.


\section{Acknowledgements}

We would like to thank Degussa, B.~Binks and A.~Schofield for providing particles. A.~Imhof is acknowledged for his \emph{miescat} software. We are grateful to J.~Arlt for his assistance with two-photon fluorescence imaging and to the EPSRC Laser Loan Pool for the loan of the ultrafast laser system. We would like to thank M.~Cates, D.~Frenkel, T.~Lapp, J.~Tavacoli, J.~Vollmer and K.~White for useful discussions. Finally, we acknowledge funding through EPSRC EP/E030173/01.

\clearpage


\thispagestyle{empty}


\clearpage

\thispagestyle{empty}


\clearpage

\thispagestyle{empty}


\clearpage

\thispagestyle{empty}


\clearpage

\thispagestyle{empty}


\clearpage

\thispagestyle{empty}


\clearpage

\clearpage

\appendix

\section{Nearly space-filling networks: an explanation involving phase diagrams}\label{SIpart:NfComp}

In the main text, we have presented the formation of polyhedral cellular networks in two different binary-liquid/particle systems en route to remixing. However, we still need to explain why we were able to fabricate nearly space-filling networks using cyclohexene-nitromethane/PMMA but not using hexane-methanol/silica. This can be accounted for by quantitatively comparing the binary-liquid phase diagrams using the lever rule\cite{Chaikin2003}
\begin{equation}\label{Eq:LeverRule}
    \phi_{\mathrm{AR}} = \left| \frac{\varphi_{\mathrm{B}}^{\mathrm{SF}} - \varphi_{\mathrm{B}}^{\mathrm{BR}}}{\varphi_{\mathrm{B}}^{\mathrm{BR}} - \varphi_{\mathrm{B}}^{\mathrm{AR}}} \right| \ .
\end{equation}
Here, $\phi_{\mathrm{AR}}$ is the volume fraction of the A-rich phase, $\varphi_{\mathrm{B}}^{\mathrm{SF}}$ the volume fraction of component B in the single-fluid phase and $\varphi_{\mathrm{B}}^{\mathrm{AR}} \left( \varphi_{\mathrm{B}}^{\mathrm{BR}} \right)$ is the volume fraction of component B in the A-rich (B-rich) phase. Note that Eq.~(\ref{Eq:LeverRule}) only holds if the global volume fraction of component B ($\varphi_{\mathrm{B}}^{\mathrm{SF}}$) does not change upon crossing the binodal.

\begin{figure*}[!h]
        \begin{center}
            \includegraphics[width=1.0\textwidth]{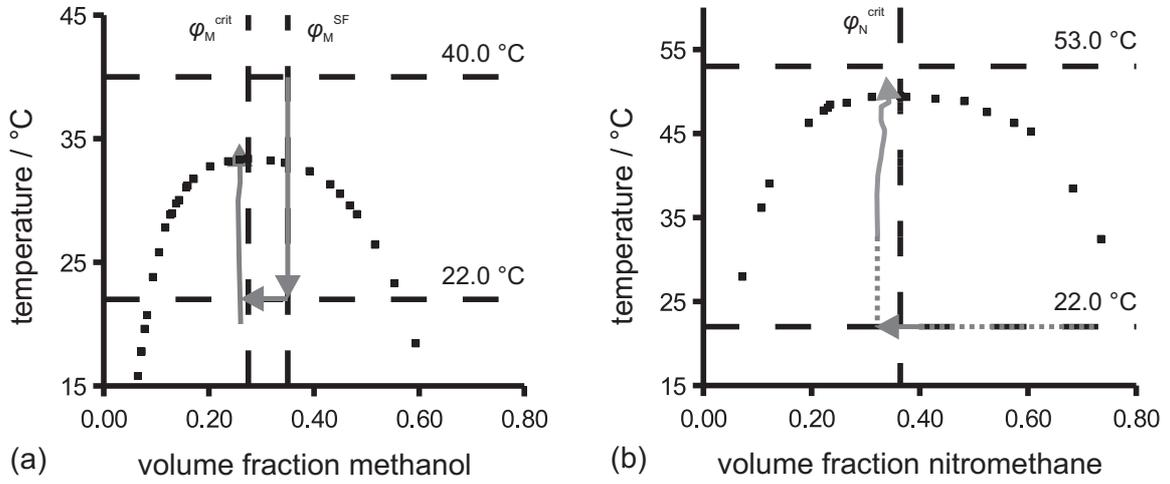}
            \caption{\small Phase diagrams for the binary mixtures (a) hexane-methanol and (b) cyclohexene-nitromethane,\cite{Hradetzky1991,Sazonov2000,PhaseDiagramConversionNote} in which $\varphi_{\mathrm{M(N)}}^{\mathrm{crit(SF)}}$ is the critical(single-fluid) volume fraction of methanol(nitromethane). The arrow pointing downward represents the quench from the single-fluid phase. The one pointing to the left depicts the system crossing the symmetry line due to creaming. Its dotted tail in panel (b) indicates that we do not know the exact initial sample composition (see Sec.~2.2 in main text). The arrow pointing upward depicts the local volume fraction of methanol(nitromethane) in the cream, calculated from the phase diagram using the lever rule (Eq.~(\ref{Eq:LeverRule})), assuming a fixed droplet packing fraction of $\phi_{\mathrm{AR}} = 64$ vol-\%.}\label{SIfig:Figure_SI_PhaseDiagrams}
        \end{center}
    \end{figure*}

The reasons for the formation of nearly space-filling networks in cyclohexene-nitromethane/PMMA are twofold. First of all, the corresponding phase diagram is more symmetric than that of hexane-methanol (Fig.~\ref{SIfig:Figure_SI_PhaseDiagrams}).\cite{Hradetzky1991,Sazonov2000,PhaseDiagramConversionNote,HandbookChemPhys20082009,Steele1996,GarciaMiaja2007} Combined with the near-critical composition of the sample (see Sec.~2.1 in main text), this results in a larger volume fraction of the dispersed phase $\phi_{\mathrm{AR}}$ at room temperature (Eq.~(\ref{Eq:LeverRule})). In other words, the creamed emulsions fill a larger proportion of the sample cell to start with.

Secondly, creaming takes the cyclohexene-nitromethane system further across the symmetry line than it does in the case of hexane-methanol. Actually, to prevent the droplets from shrinking upon heating, the creamed emulsions must have a liquid-liquid ratio on the left-hand side of the \emph{critical} composition. In the case of cyclohexene-nitromethane, this requires a droplet packing fraction in the cream $\phi_{\mathrm{droplet}}^{\mathrm{cream}}$ of only 57 vol-\% at 38 \degC{} (Fig.~\ref{SIfig:Figure_SI_PhaseDiagrams} and Eq.~(\ref{Eq:LeverRule})), whereas it must exceed 61 vol-\% at 22 \degC{} for hexane-methanol (equal $\Delta T$ to $T_{\mathrm{crit}}$). This can easily be achieved in a random packing, which has a maximum filling fraction of 64 vol-\% for monodisperse spheres. As the droplets in our emulsions are highly polydisperse (e.g.~Fig.~2e in main text), they may even pack more efficiently, up to 78 vol-\% for a polydispersity of 40\%.\cite{Schaertl1994}

As buoyancy keeps the droplets in the cream closely packed, heating the emulsion toward the single-fluid phase can be considered as a quench at constant $\phi_{\mathrm{AR}}$. This is different from cooling from the single-fluid phase, which is a quench at constant $\varphi_{\mathrm{B}}^{\mathrm{SF}}$ (Eq.~(\ref{Eq:LeverRule})). Using a fixed value of $\phi_{\mathrm{droplet}}^{\mathrm{cream}} = 64$ vol-\% as a reasonable assumption, the lever rule predicts the trajectories in Fig.~\ref{SIfig:Figure_SI_PhaseDiagrams} for the compositions of the creamed emulsions upon heating. Note that the trajectory for cyclohexene-nitromethane is further to the left of the critical-composition line than for hexane-methanol. In fact, for a fixed value of $\varphi_{\mathrm{N}}^{\mathrm{cream}} = 0.32$, corresponding to $\phi_{\mathrm{droplet}}^{\mathrm{cream}} = 64$ vol-\% at 38 \degC{}, the phase diagram predicts that the volume fraction of the cyclohexene-rich phase ($\phi_{\mathrm{AR}}$ in Eq.~(\ref{Eq:LeverRule})) should rapidly increase close to the binodal. As no new droplets are formed, the existing ones will tend to grow, i.e. the cream will expand.

\clearpage



\clearpage


\begin{thebibliography}{10}

\bibitem{BinksHorozov2008}
B.~P. Binks and T.~S. Horozov, editors.
\newblock {\em Colloidal Particles at Liquid Interfaces}.
\newblock Cambridge University Press, Cambridge (UK), 2008.

\bibitem{Clegg2007}
P.~S. Clegg, E.~M. Herzig, A.~B. Schofield, S.~U. Egelhaaf, T.~S. Horozov,
  B.~P. Binks, M.~E. Cates, and W.~C.~K. Poon.
\newblock Emulsification of partially miscible liquids using colloidal
  particles: Nonspherical and extended domain structures.
\newblock {\em Langmuir}, 23(11):5984--5994, 2007.

\bibitem{Hertlein2008}
C.~Hertlein, L.~Helden, A.~Gambassi, S.~Dietrich, and C.~Bechinger.
\newblock Direct measurement of critical casimir forces.
\newblock {\em Nature}, 451(7175):172--175, 2008.

\bibitem{Araki2008}
T.~Araki and H.~Tanaka.
\newblock Dynamic depletion attraction between colloids suspended in a
  phase-separating binary liquid mixture.
\newblock {\em J.~Phys.: Condens.~Matter}, 20(7):072101, 2008.

\bibitem{Lu2008}
X.~H. Lu, S.~G.~J. Mochrie, S.~Narayanan, A.~R. Sandy, and M.~Sprung.
\newblock How a liquid becomes a glass both on cooling and on heating.
\newblock {\em Phys.~Rev.~Lett.}, 100(4):045701, 2008.

\bibitem{Koehler1997}
R.~D. Koehler and E.~W. Kaler.
\newblock Colloidal phase transitions in aqueous nonionic surfactant solutions.
\newblock {\em Langmuir}, 13(9):2463--2470, 1997.

\bibitem{Beysens1999}
D.~Beysens and T.~Narayanan.
\newblock Wetting-induced aggregation of colloids.
\newblock {\em J.~Stat.~Phys.}, 95(5-6):997--1008, 1999.

\bibitem{Gallagher1992}
P.~D. Gallagher and J.~V. Maher.
\newblock Partitioning of polystyrene latex spheres in immiscible critical
  liquid-mixtures.
\newblock {\em Phys.~Rev.~A}, 46(4):2012--2021, 1992.

\bibitem{Loewen1995}
H.~$\mathrm{L\ddot{o}wen}$.
\newblock Solvent-induced phase-separation in colloidal fluids.
\newblock {\em Phys.~Rev.~Lett.}, 74(6):1028--1031, 1995.

\bibitem{Peng2000}
G.~W. Peng, F.~Qiu, V.~V. Ginzburg, D.~Jasnow, and A.~C. Balazs.
\newblock Forming supramolecular networks from nanoscale rods in binary,
  phase-separating mixtures.
\newblock {\em Science}, 288(5472):1802--1804, 2000.

\bibitem{Stratford2005}
K.~Stratford, R.~Adhikari, I.~Pagonabarraga, J.~C. Desplat, and M.~E. Cates.
\newblock Colloidal jamming at interfaces: A route to fluid-bicontinuous gels.
\newblock {\em Science}, 309(5744):2198--2201, 2005.

\bibitem{Herzig2007}
E.~M. Herzig, K.~A. White, A.~B. Schofield, W.~C.~K. Poon, and P.~S. Clegg.
\newblock Bicontinuous emulsions stabilized solely by colloidal particles.
\newblock {\em Nat.~Mater.}, 6:966--971, 2007.

\bibitem{Tanaka2000}
H.~Tanaka.
\newblock Viscoelastic phase separation.
\newblock {\em J.~Phys.: Condens.~Matter}, 12(15):R207--R264, 2000.

\bibitem{Abbas1997}
S.~Abbas, J.~Satherley, and R.~Penfold.
\newblock The liquid-liquid coexistence curve and the interfacial tension of
  the methanol-n-hexane system.
\newblock {\em J.~Chem.~Soc., Faraday Trans.}, 93(11):2083--2089, 1997.

\bibitem{HandbookChemPhys20082009}
David~R. Lide, editor.
\newblock {\em Handbook of Chemistry and Physics}.
\newblock CRC Press, Boca Raton (USA), $89^{\mathrm{th}}$ edition, 2008-2009.
\newblock Internet Version.

\bibitem{Tanaka1994}
H.~Tanaka, A.~J. Lovinger, and D.~D. Davis.
\newblock Pattern evolution caused by dynamic coupling between wetting and
  phase-separation in binary-liquid mixture containing glass particles.
\newblock {\em Phys.~Rev.~Lett.}, 72(16):2581--2584, 1994.

\bibitem{Chung2004}
H.~J. Chung, A.~Taubert, R.~D. Deshmukh, and R.~J. Composto.
\newblock Mobile nanoparticles and their effect on phase separation dynamics in
  thin-film polymer blends.
\newblock {\em Europhys.~Lett.}, 68(2):219--225, 2004.

\bibitem{Hunter2008}
T.~N. Hunter, R.~J. Pugh, G.~V. Franks, and G.~J. Jameson.
\newblock The role of particles in stabilising foams and emulsions.
\newblock {\em Adv.~Colloid Interface Sci.}, 137(2):57--81, 2008.

\bibitem{Banhart2001}
J.~Banhart.
\newblock Manufacture, characterisation and application of cellular metals and
  metal foams.
\newblock {\em Prog.~Mater.~Sci.}, 46(6):559--632, 2001.

\bibitem{Cook2008}
Catherine Cook, Woo Renee, Brenda O'Neil, McKenzie Leigh, Jason Manning, and
  Martin~Gerard Bakker.
\newblock Development of oil/water/surfactant microemulsions as templates for
  micro and nanostructured metal foams.
\newblock {\em Mater.~Res.~Soc.~Symp.~Proc.}, 1059:1059--KK10--32, 2008.

\bibitem{Hradetzky1991}
G.~Hradetzky and D.~A. Lempe.
\newblock Phase-equilibria in binary and higher systems methanol +
  hydrocarbon(s) : Part i. experimental determination of liquid-liquid
  equilibrium data and their representation using the nrtl equation.
\newblock {\em Fluid Phase Equilib.}, 69:285--301, 1991.

\bibitem{WaterNote}
Both hexane and methanol are easily contaminated with water from the air. Even
  a few volume percent is enough to raise the consolute temperature of the
  binary liquid by several degrees.\cite{Alessi1989} Contamination with water
  can be prevented by storing the hexane over molecular sieves and the methanol
  under dry nitrogen until the day of the experiment.

\bibitem{Alessi1989}
P.~Alessi, M.~Fermeglia, and I.~Kikic.
\newblock Liquid-liquid equilibrium of cyclohexane-n-hexane-methanol mixtures:
  Effect of water content.
\newblock {\em J.~Chem.~Eng.~Data}, 34(2):236--240, 1989.

\bibitem{IsohexaneNote}
Using iso-hexane instead of n-hexane causes a decrease in the consolute
  temperature of the hexane-methanol mixture of a few degrees.

\bibitem{Sazonov2000}
V.~P. Sazonov, K.~N. Marsh, and G.~T. Hefter.
\newblock Iupac-nist solubility data series 71. nitromethane with water or
  organic solvents: Binary systems.
\newblock {\em J.~Phys.~Chem.~Ref.~Data}, 29(5):1165--1354, 2000.

\bibitem{Bosma2002}
G.~Bosma, C.~Pathmamanoharan, E.~H.~A. de~Hoog, W.~K. Kegel, A.~van Blaaderen,
  and H.~N.~W. Lekkerkerker.
\newblock Preparation of monodisperse, fluorescent pmma-latex colloids by
  dispersion polymerization.
\newblock {\em J.~Colloid Interface Sci.}, 245(2):292--300, 2002.

\bibitem{ImageJ138x}
ImageJ 1.38x/1.41o, Wayne Rasband, NIH (USA), http://rsb.info.nih.gov/ij/.

\bibitem{BlankNote}
Unfortunately, confocal reflection images also reveal the edges of isolated
  droplets in samples without any silica. However, these are not as fuzzy as
  the ones in samples with silica particles and they appear to consist of
  several interference fringes.

\bibitem{AveragingNote}
Fig.~\ref{fig:Figure_TwoPhoton}a was extracted from an $xyz$-series in which
  scans were not averaged, while each frame in the $xyz$-series corresponding
  to Figs.~\ref{fig:Figure_TwoPhoton}c and \ref{fig:Figure_TwoPhoton}e was an
  average of two and three scans, respectively. However,
  Figs.~\ref{fig:Figure_TwoPhoton}d and \ref{fig:Figure_TwoPhoton}f were only
  averaged $10 \times$ whereas Fig.~\ref{fig:Figure_TwoPhoton}b was averaged
  $15 \times$.

\bibitem{Sanz2009}
E.~Sanz, K.~A. White, P.~S. Clegg, and M.~E. Cates.
\newblock Colloidal gels assembled via a temporary interfacial scaffold.
\newblock {\em Phys.~Rev.~Lett.}, 103(25):255502, 2009.

\bibitem{Meeker2000}
S.~P. Meeker, W.~C.~K. Poon, J.~Crain, and E.~M. Terentjev.
\newblock Colloid-liquid-crystal composites: An unusual soft solid.
\newblock {\em Phys.~Rev.~E}, 61(6):R6083--R6086, 2000.

\bibitem{Vollmer2005}
D.~Vollmer, G.~Hinze, B.~Ullrich, W.~C.~K. Poon, M.~E. Cates, and A.~B.
  Schofield.
\newblock Formation of self-supporting reversible cellular networks in
  suspensions of colloids and liquid crystals.
\newblock {\em Langmuir}, 21(11):4921--4930, 2005.

\bibitem{Colard2009}
C.~A.~L. Colard, R.~A. Cave, N.~Grossiord, J.~A. Covington, and S.~A.~F. Bon.
\newblock Conducting nanocomposite polymer foams from ice-crystal-templated
  assembly of mixtures of colloids.
\newblock {\em Adv.~Mater.}, 21(28):2894--2898, 2009.

\bibitem{Chaikin2003}
P.~M. Chaikin and T.~C. Lubensky.
\newblock {\em Principles of condensed matter physics}.
\newblock Cambridge University Press, Cambridge (UK), 2003.

\bibitem{PhaseDiagramConversionNote}
Mole fractions were converted into volume fractions using linear fits to the
  temperature-dependent densities of hexane,\cite{HandbookChemPhys20082009}
  methanol,\cite{HandbookChemPhys20082009} cyclohexene,\cite{Steele1996} and
  nitromethane\cite{GarciaMiaja2007}.

\bibitem{Steele1996}
W.~V. Steele, R.~D. Chirico, S.~E. Knipmeyer, A.~Nguyen, N.~K. Smith, and I.~R.
  Tasker.
\newblock Thermodynamic properties and ideal-gas enthalpies of formation for
  cyclohexene, phthalan (2,5-dihydrobenzo-3,4-furan), isoxazole, octylamine,
  dioctylamine, trioctylamine, phenyl isocyanate, and
  1,4,5,6-tetrahydropyrimidine.
\newblock {\em J.~Chem.~Eng.~Data}, 41(6):1269--1284, 1996.

\bibitem{GarciaMiaja2007}
G.~Garcia-Miaja, J.~Troncoso, and L.~Romani.
\newblock Density and heat capacity as a function of temperature for binary
  mixtures of 1-butyl-3-methylpyridinium tetrafluoroborate plus water, plus
  ethanol, and plus nitromethane.
\newblock {\em J.~Chem.~Eng.~Data}, 52:2261--2265, 2007.

\bibitem{Schaertl1994}
W.~Schaertl and H.~Sillescu.
\newblock Brownian dynamics of polydisperse colloidal hard-spheres -
  equilibrium structures and random close packings.
\newblock {\em J.~Stat.~Phys.}, 77(5-6):1007--1025, 1994.

\end{thebibliography}
\end{document}